\documentclass[aoas,nameyear,dvips]{arximspdf}
\usepackage{dcolumn}
\usepackage{graphicx}

\doi{10.1214/10-AOAS339}
\volume{4}
\issue{3}
\pubyear{2010}
\firstpage{1517}
\lastpage{1532}

\makeatletter
\newcolumntype{d}[1]{D{.}{.}{#1}}
\makeatother

\begin{document}
\begin{frontmatter}

\title{An approach for jointly modeling multivariate
longitudinal measurements and discrete time-to-event data\thanksref{T1}}
\thankstext{T1}{Supported by the Intramural Research Program of the National Institutes of Health,
\textit{Eunice Kennedy Shriver} National Institute of Child Health and Human Development.}
\runtitle{Multivariate longitudinal and time-to-event data}

\begin{aug}
\author[A]{\fnms{Paul S.} \snm{Albert}\corref{}\ead[label=e1]{albertp@mail.nih.gov}}
\and
\author[B]{\fnms{Joanna H.} \snm{Shih}}

\runauthor{P. S. Albert and J. H. Shih}
\affiliation{Eunice Kennedy Shriver National Institute of Child Health and
Human Development and National Cancer Institute}
\address[A]{Eunice Kennedy Shriver\\
\quad National Institute of Child Health\\
\quad and Human Development\\
National Institutes of Health\\
Bethesda, Maryland 20892\\
USA\\
\printead{e1}} 
\address[B]{Biometric Research Branch\\
Division of Cancer Treatment and Diagnosis\\
National Cancer Institute\\
Bethesda, Maryland 20892\\
USA}
\end{aug}

\received{\smonth{8} \syear{2009}}
\revised{\smonth{2} \syear{2010}}

\begin{abstract}
In many medical studies, patients are followed longitudinally and
interest is on assessing the relationship between longitudinal
measurements and time to an event.  Recently, various authors have
proposed joint modeling approaches for longitudinal and
time-to-event data for a single longitudinal variable.  These joint
modeling approaches become intractable with even a few longitudinal
variables.  In this paper we propose a regression calibration
approach for jointly modeling multiple longitudinal measurements and
discrete time-to-event data. Ideally, a two-stage modeling approach
could be applied in which the multiple longitudinal measurements are
modeled in the first stage and the longitudinal model is  related to
the time-to-event data in the second stage.  Biased parameter
estimation due to informative dropout makes this direct two-stage
modeling approach problematic. We propose a regression calibration
approach which appropriately accounts for informative dropout.  We
approximate the conditional distribution of the multiple
longitudinal measurements given the event time by modeling all
pairwise combinations of the longitudinal measurements using a
bivariate linear mixed model which conditions on the event time.
Complete data are then simulated based on estimates from these
pairwise conditional models, and regression calibration is used to
estimate the relationship between longitudinal data and
time-to-event data using the complete data. We show that this
approach performs well in estimating the relationship between
multivariate longitudinal measurements and the time-to-event data
and in  estimating the parameters of the multiple longitudinal
process subject to informative dropout. We illustrate this
methodology with simulations and with an analysis of  primary
biliary cirrhosis (PBC) data.
\end{abstract}

\begin{keyword}
\kwd{Joint models}
\kwd{shared random parameter models}
\kwd{informative dropout}
\kwd{regression calibration}.
\end{keyword}

\end{frontmatter}

\section{Introduction}\label{s1}

Recently, many studies collect longitudinal data on a panel of
biomarkers, and interest is on assessing the relationship between
these biomarkers and time to an event. For example, \citet{allen} examined the relationship between
 five  longitudinally collected cytokines measured from serum plasma and
survival. Interest focused on whether the values of these
multiple cytokines are associated with survival.
In another example, patients with primary biliary cirrhosis are followed longitudinally and interest is on examining whether multiple longitudinally  biomarkers are prognostic for a poor clinical outcome. Important  features in studies of this type are  that there may be a relatively large number of biomarkers and that these  biomarkers are subject to sizable measurement error due to
laboratory error and biological variation.

Various authors have proposed joint modeling approaches for a single
longitudinal measurement and time-to-event data [\citet{tsiatis1995}; \citet{wulfsohn1997}; \citet{tsiatis2004}; \citet{henderson}, among others]. There is also limited work
on joint models for a few longitudinal measurements and
time-to-event data [Xu and Zeger (\citeyear{xu2001a}, \citeyear{xu2001b}); \citet{huang}; \citet{song}; \citet{ibrahim};
\citet{brown}; \citet{chi}]. However, these methods
are difficult to implement when the number of longitudinal
biomarkers is large since most of these approaches  require
integrating over the vector of all random effects to evaluate  the
joint likelihood of the multivariate longitudinal and time-to-event
data. This paper proposes an approach for jointly modeling
multivariate longitudinal and discrete time-to-event data which
easily accommodates many longitudinal biomarkers.

\citet{fieuws2005} and \citet{fieuws2007} have  proposed
an approach for modeling multivariate longitudinal data whereby all possible pairs of longitudinal data are
 separately modeled and are then combined in a final step. We use a similar approach along with
a recent regression calibration approach for jointly modeling a single series of
longitudinal measurements and time-to-event data [\citet{albert}] to implement the  joint modeling approach
 proposed in this paper. Recently, \citet{fieuws2008} have proposed a discriminant analysis based approach for using
 multivariate longitudinal profiles to predict renal graft failure. At the end of their discussion,
 they mention that a more elegant approach, which has not yet been developed,
  would involve a  joint model for the many longitudinal profiles and time-to-event data.
  This paper presents such an approach.

We describe the approach in Section \ref{s2}.  We show the advantages of this approach using simulation in Section \ref{s3}. We
illustrate the methodology with an analysis of primary biliary cirrhosis data (PBC) in which we simultaneously examine
 the relationship between multiple longitudinal biomarker in Section \ref{s4}. A discussion follows in Section \ref{s5}.

\section{Modeling approach}\label{s2}
Define $T_i$ to be a discrete event-time which can take on discrete
values $t_j$, $j=1,2,\ldots,J$, and $Y_{ij}$ to be a binary indictor
of whether the $i$th patient is dead at time $t_j$. Then
$J_i=\sum_{j=1}^J (1-Y_{ij})=J-Y_{i\cdot}$, where
$Y_{i\cdot}=\sum_{j=1}^J Y_{ij}$ indicates the number of follow-up
measurements before the event or the
 end of follow-up at time $t_J$.
 Longitudinal measurements are measured at times $t_1,t_2,\ldots,t_{J_i}$.
Denote $\mathbf{X}_{1i}=(X_{1i1},X_{1i2},\ldots,X_{1iJ_i})'$, $\mathbf{X}_{2i}=(X_{2i1},X_{2i2},\ldots,X_{2iJ_i})',
\ldots,\mathbf{X}_{Pi}=(X_{Pi1},X_{Pi2},\ldots,X_{PiJ_i})'$ as the $P$
biomarkers measured repeatedly at $j=1,2,\ldots,J_i$ time points.
Further, define
$\mathbf{X}_{pi}^*=(X_{pi1}^*,X_{pi2}^*,\ldots,X_{piJ_i}^*)'$ as the
longitudinal measurements without measurement error  for the $p$th
biomarker and
$\mathbf{X}_i^*=(\mathbf{X}_{1i}^*,\mathbf{X}_{2i}^*,\ldots,\mathbf{X}_{Pi}^*)$.
We consider a joint model for multivariate longitudinal and discrete
time-to-event data in which the discrete event time distribution is
modeled as a linear function of previous true values of the
biomarkers without measurement error on the probit scale.
Specifically,
\begin{equation}\label{1f}
P\bigl(Y_{ij}=1|Y_{i(j-1)}=0;\mathbf{X}_i^*\bigr)=\Phi\Biggl(\alpha_{0j}+\sum_{p=1}^P\alpha_p X^*_{pi(j-1)}\Biggr),
\end{equation}
where $i=1,2,\ldots,I$, $j=2,3,\ldots,J_i$, $Y_{i1}$ is
taken as $0$, $\alpha_{0j}$ governs the baseline discrete event time
distribution and $\alpha_p$ measures the effect of the $p$th
biomarker ($p=1,2,\ldots,P$) at time $t_{j-1}$  on survival at time
$t_j$. Specifically, (\ref{1f}) allows for examining the effect of multiple
``true'' biomarker values at time $j-1$ on the probability of an
event between the $(j-1)$th and $j$th time point.

The longitudinal data is modeled assuming that the fixed and random effect
trajectories are linear. Specifically,
the multivariate longitudinal biomarkers can be modeled as
\begin{equation}\label{2f}
X_{pij}=X^*_{pij}+\varepsilon_{pij},
\end{equation}
where
\begin{equation}\label{3f}
X^*_{pij}=\beta_{p0}+\beta_{p1}t_j+\gamma_{pi0}+\gamma_{pi1}t_j,
\end{equation}
where $\beta_{p0}$ and $\beta_{p1}$ are the fixed effect
intercept and slope for the $p$th biomarker, and  $\gamma_{pi0}$ and
$\gamma_{pi1}$ are the random effect intercept and slope for the
$p$th biomarker on the $i$th individual. Denote
$\bolds{\beta}=(\beta_{10}, \beta_{11}, \beta_{20},
\beta_{21},\ldots,\beta_{P0},\break\beta_{P1})'$ and $\bolds{\gamma}_i
=(\gamma_{1i0}, \gamma_{1i1},\gamma_{2i0},\gamma_{2i1},\ldots,
\gamma_{Pi0}, \gamma_{Pi1})'$. We assume that $\bolds{\gamma}_{i}$ is
normally distributed with mean $\mathbf{0}$ and variance
$\bolds{\Sigma}_{\bolds{\gamma}}$, where
$\bolds{\Sigma}_{\bolds{\gamma}}$ is a $2P$ by $2P$ dimensional
variance matrix, and $\varepsilon_{pij}$ are independent error terms
which are assumed to be normally distributed with mean 0 and
variance $\sigma_{p\varepsilon}^2$ ($p=1,2,\ldots,P$).

Alternative to (\ref{1f}), where the probability of an event  over an
interval depends on the true biomarker values at the beginning of
the interval, the event-time process could be adapted to depend on
the random effects of the multivariate longitudinal process [e.g.,
$\gamma_{pi1}$ can replace $X_{pi(j-1)}^*$ in (\ref{1f})].

\subsection{Difficulty in joint estimation}\label{s21}

Conceptually, model (\ref{1f})--(\ref{2f}) can be  estimated by maximizing the likelihood
 \begin{eqnarray}\label{4f}
 L&=&\prod_{i=1}^I
   \int_{\bolds{\gamma}_i}\cdots\int\Biggl\{\prod_{p=1}^P
   h(\mathbf{X}_{pi}|\gamma_{pi0},\gamma_{pi1})\Biggr\}\nonumber\\[-8pt]\\[-8pt]
    &&\hspace*{56pt}{}\times\Biggl\{\prod_{j=2}^{J_i}(1-r_{ij})\Biggr\}\bigl(r_{i(J_i+1)}\bigr)^{J_i<J}f(\bolds{\gamma}_i)\,d\bolds{\gamma}_i,\nonumber
       \end{eqnarray}
 where $r_{ij}=P(Y_{ij}=1|Y_{i(j-1)}=0)$, $h (\mathbf{X}_{pi}|\gamma_{pi0},\gamma_{pi1})$ is the product of
 $J_i$ univariate normal density functions each with mean $X_{pij}^*$ and variance
 $\sigma_{p\varepsilon}^2$, and $f(\bolds{\gamma})$ is a multivariate normal density with mean zero and variance
 $\bolds{\Sigma}_{\bolds{\gamma}}$.  When $P=1$, (\ref{4f}) can be maximized
  by numerical integration
 techniques such as a simple trapezoidal rule or Gaussian quadrature [\citet{abramowitz}]. However, these methods are not feasible for even a few longitudinal biomarkers. Alternative
 Monte Carlo methods such as Monte Carlo EM [\citet{wei}] are possible, but these methods do not perform well for even moderately high dimensional random effects (say, $P>2$).
 In the next subsection we develop an alternative approach which is easy  to implement with a large number of longitudinal biomarkers.

 \subsection{Estimation}\label{s22}

   We propose a two stage regression calibration approach for estimation, which
   can be described as follows.
 In the first stage, multivariate linear mixed models can be used to model the longitudinal data. In the second stage,
  the time-to-event model is estimated by replacing the random effects with corresponding empirical Bayes estimates.
 There are three  problems with directly
 applying this approach. First, estimation in the first stage is complicated by the fact that simply fitting
  multivariate linear mixed models results in
 bias due to informative dropout; this is demonstrated by \citet{albert} for the the case of $P=1$. Second,
 as described in Section \ref{s21}, parameter estimation for multivariate linear mixed models can be computationally difficult
  when the number of longitudinal measurements ($P$)  is even moderately large.  Third, calibration error in the
 empirical Bayes estimation needs to be accounted for in the time-to-event model. The proposed approach will deal with all three of these problems.

 The bias from informative dropout is a result of differential follow-up whereby the longitudinal process is related to the length of follow-up.  That is, in (\ref{2f})--(\ref{3f}), patients with large values of $X_{pij}^*$ are more likely to have an early event when $\alpha_p>0$ for
  $p=1,2,\ldots,P$. There would be no bias if all $J$ follow-up measurements were observed on all patients. As proposed by \citet{albert} for univariate longitudinal data, we can avoid this bias by  generating complete data from the conditional distribution of
 $\mathbf{X}_i=(\mathbf{X}_{1i},\mathbf{X}_{2i},\ldots,\mathbf{X}_{Pi})$ given $T_i$, denoted as $\mathbf{X}_i|T_i$.
 Since $\mathbf{X}_i|T_i$ under model (\ref{2f})--(\ref{3f}) does not have a tractable form, we propose a simple approximation for this conditional distribution.  Under model (\ref{2f})--(\ref{3f}), the distribution of $\mathbf{X}_i|T_i$ can be expressed as
 \begin{equation}\label{5f}
 P(\mathbf{X}_i|T_i)=\int h(\mathbf{X}_i|\bolds{\gamma}_i,T_i)g(\bolds{\gamma}_i|T_i)\,d\bolds{\gamma}_i.
 \end{equation}
 Since $T_i$ and the values of $\mathbf{X}_i$ are conditional independent given
 $\bolds{\gamma}_i$,\break
 $h(\mathbf{X}_i|\bolds{\gamma}_i,T_i)=h(\mathbf{X}_i|\bolds{\gamma}_i)$, where
 $h(\mathbf{X}_i|\bolds{\gamma}_i)=\prod_{p=1}^P h(\mathbf{X}_{pi}|\gamma_{pi0},\gamma_{pi1})$.
  The distribution of $\mathbf{X}_i|T_i$
 can be expressed as a multivariate linear mixed model if we approximate
 $g(\bolds{\gamma}_i|T_i)$ by a normal distribution. Under the assumption that
 $g(\bolds{\gamma}_i|T_i)$ is normally distributed with mean
 $\bolds{\mu}_{T_i}=(\mu_{01T_i},\mu_{11T_i},\mu_{02T_i},\mu_{12T_i},\break\ldots,\mu_{0PT_i},\mu_{1PT_i})'$
  and variance $\bolds{\Sigma}_{\bolds{\gamma}T_i}^*$, and by rearranging mean structure parameters in the integrand of (\ref{5f}) so that the random effects have mean zero,
 $\mathbf{X}_i|T_i$ corresponds to the following multivariate linear mixed model:
 \begin{equation}\label{6f}
 X_{pij}|(T_i,\gamma^*_{ip0T_i},\gamma^*_{ip1T_i})=\beta_{p0T_i}^*+\beta_{p1T_i}^*t_j+
 \gamma^*_{ip0T_i}+\gamma^*_{ip1T_i}t_j+\varepsilon^*_{pij},
 \end{equation}
where $i=1,2,\ldots,I$, $j=1,2,\ldots,J_i$, and $p=1,2,\ldots,P$. The parameters $\beta_{p0T_i}^*$ and $\beta_{p1T_i}^*$ are intercept and slope parameters for the $p$th longitudinal measurement and for patients who have an event time at time $T_i$ or who are censored at time $t_J$.
 In addition, the associated random effects $\bolds{\gamma}_{iT_i}^*=
 (\gamma^*_{i10T_i},\gamma^*_{i11T_i},\gamma^*_{i20T_i},\gamma^*_{i21T_i},\ldots,\gamma^*_{iP0T_i},\gamma^*_{iP1T_i})'$
 are multivariate normal with mean $\mathbf{0}$ and variance $\bolds{\Sigma}_{\bolds{\gamma} T_i}^*$, and the residuals $\varepsilon_{pij}^*$ are assumed to have an independent normal distribution with mean zero and variance $\sigma_{\varepsilon p}^{*2}$.
 Thus, this conditional model involves estimating separate fixed effect intercept and slope parameters for each potential event-time and for subjects who are censored at
 time $t_J$. Likewise, separate random effects distributions are estimated for each of these discrete time points. For example, the intercept and slope fixed-effect parameters for the $p$th biomarker  for those\vspace{1pt} patients who have an event at time $T_i=t_3$ is $\beta_{p0t_3}^*$ and
 $\beta_{p1t_3}^*$, respectively.
 Further, the intercept and slope random effects for all $P$ biomarkers on those
 patients who have an event at time $T_i=t_3,\bolds{\gamma}^*_{it_3} $, is multivariate
 normal with mean $\mathbf{0}$ and variance $\bolds{\Sigma}_{\bolds{\gamma}t_3}^*$.
 A similar approximation has been proposed by \citet{albert} for univariate
 longitudinal data $(P=1)$.

Recall that by generating complete data from (\ref{6f}) we are able to
avoid the bias due to informative dropout. However, when $P$ is
large,  direct estimation of model (\ref{6f}) is difficult since the number
of elements in $\bolds{\Sigma}^*_{\bolds{\gamma}T_i}$ grows
quadratically  with $P$. For example, the dimension of the variance
matrix $\bolds{\Sigma}^*_{\bolds{\gamma}T_i}$ is $2P$ by $2P$ for $P$
longitudinal biomarkers.
 \citet{fieuws2005} proposed estimating the parameters of multivariate  linear mixed models by formulating bivariate linear mixed models
on all possible pairwise combinations of longitudinal measurements.
In the simplest approach, they proposed fitting bivariate linear
mixed models on all $P\choose 2$ combinations of longitudinal
biomarkers and averaging ``overlapping'' or duplicate  parameter
estimates. Thus, we estimate the parameters in the fully specified
model (\ref{6f}) by fitting $P \choose 2$ bivariate longitudinal models
that only include pairs of longitudinal markers. \citet{fieuws2005} demonstrated with simulations that there is little efficiency
loss using their approach relative to a full maximum-likelihood
based approach.
 Fitting these
bivariate models is computationally feasible since only four
correlated random effects are contained in each model. (That is,
$\bolds{\Sigma}^*_{\bolds{\gamma}T_i}$ is a four by four-dimensional
matrix for each discrete event-time $T_i$.) Duplicate estimates of
fixed effects and random-effect variances from all pairwise
bivariate models are averaged to obtain final parameter estimates of
the fully specified model (\ref{6f}). For example, when $P=4$ there are
$(P-1)=3$ estimates of $\beta^*_{p0T_i}$, $\beta^* _{p1T_i}$,
$\sigma_{\varepsilon p}^{*2}$ for  the $p$th longitudinal biomarker
that need to be averaged.

Model (\ref{6f}) is then used to construct complete longitudinal pseudo data sets
which  in turn are used to estimate the mean of the posterior distribution of an individual's random effects given the data.
Specifically,
multiple complete longitudinal data sets can be constructed by simulating $X_{pij}$ values from the approximation to the distribution of $\mathbf{X}_i|T_i$ given by (\ref{6f}) where the parameters are replaced by their estimated values.
Since the simulated data sets have complete follow-up on each individual,
the bias in estimating the posterior mean of $\bolds{\gamma}_i$ caused
by informative dropout will be much reduced.

The posterior mean of  distribution $\bolds{\gamma}_i$ given the data
can be estimated by fitting (\ref{2f})--(\ref{3f}) to the generated complete
longitudinal pseudo data. However, similar to fitting the
conditional model (\ref{6f}), fitting model (\ref{2f})--(\ref{3f}) is difficult due to the
high dimension of $\bolds{\Sigma}_{\bolds{\gamma}}$. Thus, we again
use the pairwise estimation approach of \citet{fieuws2005},
whereby we estimate  the parameters of (\ref{2f})--(\ref{3f}) by fitting  all
pairwise bivariate models and averaging duplicate parameter
estimates to obtain final parameter estimates.  For each generated
complete longitudinal pseudo data set, the estimate of the posterior
mean, denoted as $\widehat{\bolds{\gamma}}_i=(\widehat{\gamma}_{1i0},\widehat{\gamma}_{1i1},\ldots,
\widehat{\gamma}_{Pi0},\widehat{\gamma}_{Pi1})'$, can be calculated as
\begin{equation}\label{7f}
\widehat{\bolds{\gamma}}_i=\bolds{\Sigma}_{\bolds{\gamma}}\mathbf{Z}_i' \mathbf{V}_i^{-1} (\mathbf{X}_i-\mathbf{Z}_i\widehat{\bolds{\beta}}),
\end{equation}
where $\mathbf{Z}_i$ is a $PJ\times 2P$ design matrix
corresponding to  the fixed and random effects in (\ref{2f})--(\ref{3f}), where
$\mathbf{Z}_i=\operatorname{diag}(\underbrace{A',A',\ldots,A')}_{P\ \mathrm{times}}$,
$A=\left({1\atop t_1}\enskip{1\atop t_2}\enskip{\cdots\atop\cdots}\enskip{1\atop t_J}\right)$, and $\mathbf{V}_i$ is the variance of
$\mathbf{X}_i$. Estimates of ${X}^*_{pij}$, denoted as
$\widehat{X}_{pij}^*$, are obtained by substituting
$(\widehat{\beta}_{p0},\widehat{\beta}_{p1}, \widehat{\gamma}_{pi0},
\widehat{\gamma}_{pi1})$ for
$({\beta}_{p0},{\beta}_{p1},{\gamma}_{pi0}, {\gamma}_{pi1})$ in (\ref{3f}).

To account for the measurement error in using $\widehat{\bolds{\gamma}}_i$ as compared with using $\bolds{\gamma}_i$ in (\ref{1f}), we note that
\begin{equation}\label{8f}
\qquad P\bigl(Y_{ij}=1|Y_{i(j-1\bigr)}=0;\widehat{\mathbf{X}}^*_i)= \Phi\biggl(\frac{\alpha_{0j}+\sum_{p=1}^P\alpha_p\widehat{X}_{pij}^*}
{\sqrt{1+\operatorname{Var}\{\sum_{p=1}^P\alpha_p(\widehat{X}_{pij}^*-X_{pij}^*})\}}\biggr),
\end{equation}
where $\operatorname{Var}\{\sum_{p=1}^P\omega_p(\widehat{X}_{pi(j-1)}^*-X_{pi(j-1}^*)\}={R}_{ij}'\operatorname{Var}(\widehat{\bolds{\gamma}}_i-\bolds{\gamma}_i)R_{ij}$,
$R_{ij}=(\omega_1,\break\omega_1t_{j-1},  \omega_2, \omega_2t_{j-1},\ldots, \omega_p,\omega_pt_{j-1})$, $\operatorname{Var}(\widehat{\bolds{\gamma}}_i-\bolds{\gamma}_i)=\bolds{\Sigma}_{\bolds{\gamma}}-\bolds{\Sigma}_{\bolds{\gamma}}
\mathbf{Z}_i'\{\mathbf{V}^{-1}_i\mathbf{Z}_i-
\mathbf{V}^{-1}_i\times\mathbf{Z}_i\mathbf{Q}\mathbf{Z}_i'\mathbf{V}^{-1}_i\}\mathbf{Z}_i\bolds{\Sigma}_{\bolds{\gamma}}$,
and where $Q=(\sum_{i=1}^I\mathbf{Z}_i'\mathbf{V}^{-1}_i\mathbf{Z}_i)^{-1}$
[\citet{laird}; \citet{verbeke}].
Expression (\ref{8f}) follows from the fact that $E[\Phi(a+V)]=\Phi[(a+\mu)/\sqrt{1+\tau^2}]$, where $V\sim N(\mu,\tau^2)$.

In the second stage, $\alpha_{0j}$ ($j=1,2,\ldots,J$) and
$\alpha_{p}$ ($p=1,2,\ldots,P$) can be estimated by maximizing the
likelihood
\begin{eqnarray}\label{9f}
L&=&\prod_{i=1}^I\Biggl[\prod_{j=2}^{J_i}
\bigl\{1-P\bigl(Y_{ij}=1|Y_{i(j-1)}=0;\widehat{\mathbf{X}}^*_i\bigr)\bigr\}\Biggr]\nonumber\\[-8pt]\\[-8pt]
&&\hspace{13pt}{}\times P\bigl(Y_{i(J_i+1)}=1|Y_{iJ_i}=0;\widehat{\mathbf{X}}^*_i\bigr)^{J_i<J},\nonumber
\end{eqnarray}
where $P(Y_{ij}=1|Y_{i(j-1}=0,\widehat{\mathbf{X}}^*_i)$ is
given by (\ref{8f}). Thus, we propose the following algorithm for
estimating $\alpha_{0j}$ and $\alpha_p$ ($p=1,2,\ldots,P$):

\begin{enumerate}
\item Estimate the parameters of  model (\ref{6f}) by fitting  ${ P \choose 2}$ bivariate models to each of the pairwise combinations of longitudinal measurements and averaging
duplicate parameter estimates. The bivariate models can be fit in  R [\citet{venables}] using
code presented in \citet{doran}.

\item Simulate complete longitudinal pseudo measurements (i.e., $X_{pij}$ for $p=1,2,\ldots,P$,  $i=1,2,\ldots,I$, $j=1,2,\ldots,J$)
 from model (\ref{6f}) with model parameters estimated from step 1.

\item Estimate the parameters in model (\ref{2f})--(\ref{3f}) without regard to the event time distribution from  complete longitudinal pseudo  measurements (simulated in step 2) by fitting all possible ${ P \choose 2}$ bivariate longitudinal models and averaging  duplicate model parameter estimates.

\item Calculate $\widehat{\bolds{\gamma}}_i$  using (\ref{7f}) and
$\widehat{\mathbf{X}}^*_{pij}$ using (\ref{3f}) with $\bolds{\gamma}_i$ replaced by $\widehat{\bolds{\gamma}}_i$ and  $\bolds{\beta}$ being replaced
by $\widehat{\bolds{\beta}}$ estimated in step 3.

\item Estimate $\alpha_{0j}$ $(j=2,3,\ldots,J$) and $\alpha_p$ $(p=1,2,\ldots,P$) using (\ref{8f}) and (\ref{9f}).

\item Repeat steps 2 to 5 $M$ times and average $\widehat{\alpha}_{0j}$ and $\widehat{\alpha}_p$ to get final estimates.
\end{enumerate}

We choose $M=10$ in the simulations and data analysis since this was
shown to be sufficiently large for univariate longitudinal modeling
discussed in \citet{albert}. Asymptotic standard errors of
$\widehat{\alpha}_{0j}$ and $\widehat{\alpha}_p$ cannot be used for
inference since they fail to account for the missing data
uncertainty in our procedure. Standard errors and  95\% confidence
intervals of parameter estimates using the bootstrap [\citet{efron}] are as follows:

\begin{enumerate}
\item Construct a bootstrap sample of size $I$, by resampling
event-time and multivariate longitudinal data with replacement
($(T_i^b, \mathbf{X}_{1i}^b, \mathbf{X}_{2i}^b,\ldots,\mathbf{X}_{pi}^b)$)
from the $(T_i, \mathbf{X}_{1i}, \mathbf{X}_{2i},\ldots,\mathbf{X}_{pi})$.

\item Fit the proposed estimation procedure.

\item Iterate 500 times between steps 1 and 2. The bootstrap standard error is the sample standard deviation of the 500 bootstrap
estimates. The 95\% confidence intervals were constructed using the
percetile method (limits are 2.5 and 97.5 percentiles of the
bootstrap distribution).
\end{enumerate}

\subsection{Incorporating covariate dependence}\label{s23}
 Covariates can be incorporated in (\ref{3f}) by adding them directly into
the multivariate linear mixed model (\ref{6f}). Specifically, if
$X^*_{pij}=\mathbf{Z}_i\bolds{\eta}_p+\beta_{p0}+\beta_{p1}t_j+\gamma_{pi0}+\gamma_{pi1}t_j$,
where $\mathbf{Z}_i$ is a vector of covariates with $\bolds{\eta}_p$
being parameters for the $p$th biomarker, then
$P(\mathbf{X}_i|T_i,Z_i)=\int
h(\mathbf{X}_i|\bolds{\gamma}_i,\mathbf{Z}_i)g(\bolds{\gamma}_i|T_i)\,d\bolds{\gamma}_i$
and $\mathbf{X}_i|T_i$ can be approximated by a multivariate linear
mixed model with $\mathbf{Z}_i \bolds{\eta}_p$ being added to the right
side of (\ref{6f}). Estimation then proceeds as described in Section \ref{s22}
Although more difficult, covariates can also be incorporated into
(\ref{1f}). If
\begin{equation}\label{10f}
P\bigl(Y_{ij}=1|Y_{i(j-1)}=0,\mathbf{X}_i^*,\mathbf{Z}_i\bigr) =\Phi\Biggl(\alpha_{0j}+
\mathbf{Z}_i\bolds{\zeta}+\sum_{p=1}^P \alpha_p X^*_{pi(j-1)}\Biggr),
\end{equation}
then $P(\mathbf{X}_i|T_i, \mathbf{Z}_i)=\int
h(\mathbf{X}_i|\bolds{\gamma}_i,\mathbf{Z}_i)g(\bolds{\gamma}_i|T_i,\mathbf{Z}_i)\,d\bolds{\gamma}_i,$
and under the assumption that $g(\bolds{\gamma}_i|T_i,\mathbf{Z}_i)$ is
normally distributed with variance not depending on $\mathbf{Z}_i$
(which we found to be the case in simulations not shown), then
$\mathbf{X}_i|T_i,\mathbf{Z}_i$ can be approximated by a multivariate
linear mixed model with $\mathbf{Z}_i\bolds{\zeta}_{pT_i}^*$ added to
the right-hand side of (\ref{6f}). Extensive simulations showed that the
conditional distribution of $\bolds{\gamma}_i|T_i,\mathbf{Z}_i$ is
nearly normally distributed over a wide range of parameter values,
with slight departures from normality found when the relationship
between the longitudinal process and event-time processes is very
strong (i.e., $\alpha_p$'s are very large in magnitude). The
multivariate linear model approximation is flexible in that it
allows the regression parameters to vary with $T_i$. A more
parsimonious model would be to constrain the parameters such that
$\bolds{\zeta}_{pT_i}^*$ does not vary with $T_i$.

\section{Simulations}\label{s3}

\begin{table}
\tablewidth=300pt
\caption{Estimates of $\alpha_{0j}=\alpha_0$,
 $\alpha_1$, $\alpha_2$ and $\alpha_3$ from model (\protect\ref{1f}) with $\sigma_{1\varepsilon}=
 \sigma_{2\varepsilon}=\sigma_{3\varepsilon}=0.75$ and with $P=3$,   $J=5$, and $I=300$.
 Random effects are generated under a diagonal covariance matrix. Further, we
assume that $t_j=j$ and all individuals who are alive at $t_5=5$ are
administratively censored at that time point.
The means (standard deviations) from 500 simulations are presented}\label{t1}
\begin{tabular*}{315pt}{@{\extracolsep{\fill}}ld{2.2}d{2.4}d{2.4}d{2.4}@{}}
\hline
\textbf{Parameters}&\multicolumn{1}{c}{\textbf{True values}}&\multicolumn{1}{c}{\textbf{Truth}}&\multicolumn{1}{c}{\textbf{Proposed}}&\multicolumn{1}{c@{}}{\textbf{Observed}}\\
\hline
$\alpha_0$&-1.75&-1.77&-1.76&-1.37\\
        & & (0.115)&(0.180)&(0.089)\\
$\alpha_1$&0.40&0.408&0.405&0.221\\
       & &(0.060)&(0.089)&(0.042)\\
$\alpha_2$&0&0.00&0.00&0.001\\
       & &(0.058)&(0.077)&(0.042)\\
$\alpha_3$&0.40&0.405&0.400&0.219\\
       & &(0.062)&(0.092)&(0.043)\\
$\beta_{10}$& 1.0&   &1.00     &       \\
          &    &   &  (0.058)   &        \\
$\beta_{11}$& 0&   &   0.03  &       \\
          &    &   &  (0.068)   &        \\
$\beta_{20}$& 0.5&   &  0.50   &       \\
          &    &   &   (0.050)  &        \\
$\beta_{21}$& 0&   &    -0.01 &       \\
          &    &   &   (0.067)  &        \\
$\beta_{30}$& 1&   &   1.00  &       \\
          &    &   &   (0.052)  &        \\
$\beta_{31}$& 0&   &   0.023  &       \\
          &    &   &   (0.065)  &        \\
$\sigma^2_{b10}$& 0.25&   &  0.277   &       \\
          &    &   &   (0.061)  &        \\
$\sigma^2_{b11}$& 0.25 &  & 0.296    &       \\
          &    &   &  (0.087)   &        \\
$\sigma^2_{b20}$& 0.25&   &  0.268   &       \\
          &    &   & (0.062)    &        \\
$\sigma^2_{b21}$& 0.25&   & 0.286    &       \\
          &    &   &   (0.087)  &        \\
$\sigma^2_{b30}$& 0.25&   & 0.270    &       \\
          &    &   &  (0.059)   &        \\
$\sigma^2_{b31}$& 0.25&   &  0.294   &       \\
          &    &   &   (0.083)  &        \\
\hline
\end{tabular*}
\end{table}

We conduct a simulation study to examine the statistical properties of
the proposed approach. The approach is examined for the situation where
$P=3$, $I=300$, and $\sigma^2_{p\varepsilon}=0.75$ for $p=1,2$ and 3.
The remaining parameters are presented in Table \ref{t1}. We compare the proposed approach
with $M=10$ with a model where the  $X^*_{pij}$'s are assumed to be known (true model) and with a model
where the observed $X_{pij}$'s are used in (\ref{1f}) (observed model).
 Although the  true values are never actually observed in practice, we examine
the true model as a benchmark in comparing the other models.
  Table \ref{t1}  shows that the proposed approach results in nearly unbiased estimates of $\alpha_1$, $\alpha_2$ and $\alpha_3$, whereas the model which uses
the observed observations (which are subject to measurement error) has severe bias for estimating $\alpha_1$ and $\alpha_3$ (the two parameters which are nonzero).
Although the variability
 of parameter estimates is larger for the proposed approach as compared with the
 observed approach, the root mean squared errors are substantially smaller for $\alpha_1$ and $\alpha_3$ for the proposed approach.
 For example, the root mean squared errors for $\alpha_1$ is 0.089 for the proposed approach and
  0.184 for the observed model.
  Results when the ``true'' values for the markers (markers without
  measurement errors) are assumed known are also presented in Table
  \ref{t1} (column labeled Truth). As expected, estimates are unbiased for
  this gold standard case. A comparison of the  gold standard case with
  the proposed approach shows efficiency loss. For example,
the relative efficiency for estimating $\alpha_1$, $\alpha_2$ and
$\alpha_3$ with the proposed approach versus the gold standard is
0.45 $[(0.06/0.089)^2]$, 0.57 and 0.24, respectively. We conducted
additional simulations with different parameter values. In all cases
tried, the mean squared errors were substantially smaller using the
proposed approach as compared with using the observed values (data
not shown).

Table \ref{t1} also presents the average estimated intercept and slope for each of the three longitudinal
 biomarkers. The results show that estimates of these fixed effects are nearly unbiased for the proposed
 approach.

\section{Example}\label{s4}

We examine the effect of multiple longitudinal biomarkers on the
short-term prognosis for patients with primary biliary cirrhosis
(PBC) using the PBC study conducted at the Mayo Clinic from 1974 to
1984 [\citet{murtaugh}]. PBC is a chronic disease characterized
by inflammatory destruction of the small bile ducts within the
liver, which eventually leads to cirrhosis of the liver, followed by
death. Various biomarkers such as biliribin, prothrombin time and
albumin were collected longitudinally, and interest is on examining
whether these biomarkers relate to the natural history of disease.
Of major interest was whether these biomarkers are prognostic for
transplantation-free survival (time to either transplantation or
death). A total of 312  patients had a baseline measurement and were
followed longitudinally at 6 months and at yearly intervals
thereafter.

\begin{figure}

\includegraphics{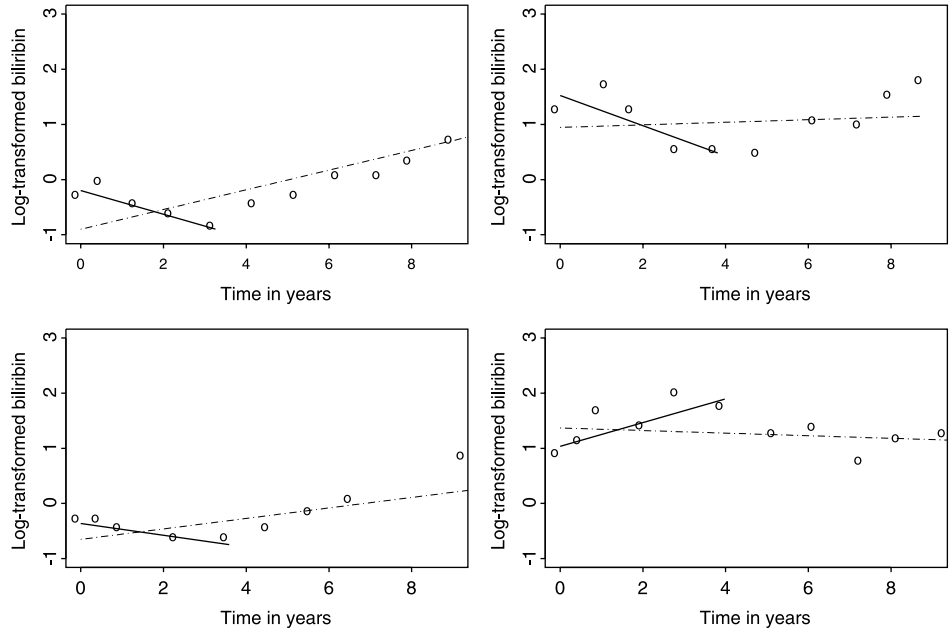}

  \caption{Plot of log-transformed biliribin values versus
follow-up time for 4 patients. Each plot shows an example of
complete follow-up in which the changes over time appear to be
linear over the first 4 years of follow-up and nonlinear over the
entire length of follow-up.  For each panel, the solid line is a
least-squares regression line for the first four years of follow-up,
while the dashed line is a corresponding line using all the
follow-up data.}\label{f1}
\end{figure}

For our application, we focused on the first 4 years of follow-up
for a number of reasons. First, individual changes in the biomarkers
appeared to be close to linear over this time period. Figure \ref{f1} shows
4 examples of complete follow-up in which the changes in
log-transformed biliribin appear to be linear over the first 4 years
of follow-up and not very linear over the whole range of follow-up.
For each panel, the solid line is a least-squares regression line
for the first four years of follow-up, while the dashed line is a
corresponding line using all the follow-up data. The patterns over
the whole follow-up period are nonlinear curves and are not
systematic over subjects, and therefore not easily characterized by
a simple nonlinear mixed model. The reason why the linear
assumption is reasonable over the shorter time interval is that,
even though the curves are nonlinear, they can adequately be
approximated as linear functions over a short time interval (i.e., a~
nonlinear function can be locally approximated by a linear
function). Second, the methodology makes the assumption that the
effect of the biomarkers on prognosis is constant over the follow-up
period [i.e., $\alpha_p$ parameters in (\ref{1f}) do not vary over time].
This assumption is more reasonable over the shorter 4 year interval
rather  than the entire follow-up period.

\begin{table}
\tablewidth=300pt
\caption{The effect of log-transformed biliribin, prothrombin time
and albumin on transplantation-free survival. The analysis is based
on fitting model (\protect\ref{1f}) with $X_{i(j-1)}$ replacing $X_{i(j-1)}^*$.
Time-to-event is modeled as a discrete time process with possible
event times at 0.5, 1, 2, 3 and 4 years, where $\alpha_{02}$
reflects the baseline discrete-time event distribution for the
intervals 0 to 0.5 and 0.5 to 1 years.  The subsequent yearly
intervals are characterized by $\alpha_{03}$, $\alpha_{04}$,
$\alpha_{05}$ and $\alpha_{06}$.  95\%  confidence intervals
 were estimated using the bootstrap with the percentile method (500 bootstrap samples)}\label{t2}
\begin{tabular*}{300pt}{@{\extracolsep{\fill}}ld{2.2}d{2.12}@{}}
\hline
\textbf{Parameters}&\multicolumn{1}{c}{\textbf{Estimate}}&\multicolumn{1}{c@{}}{\textbf{95\% confidence interval}}\\
\hline
$\alpha_{02}$&-3.71&-8.10\mbox{ to }{-}1.01\\
$\alpha_{03}$&-4.01&-8.51\mbox{ to }{-}1.14\\
$\alpha_{04}$&-3.37&-7.73\mbox{ to }{-}0.50\\
$\alpha_{05}$&-3.43&-7.91\mbox{ to }{-}0.39\\
$\alpha_{06}$&-3.39&-7.78\mbox{ to }{-}0.53\\
log Biliribin&0.58&0.45\mbox{ to }0.74\\
log Albumin&-2.57&-3.76\mbox{ to }{-}1.56\\
log Proth&1.86&0.79\mbox{ to } 3.53\\
\hline
\end{tabular*}
\end{table}

 Table \ref{t2} shows parameter estimates from fitting model
 (\ref{1f}) with the observed data as covariates instead of the  true values.
 Although both standard errors and 95\% confidence intervals were estimated using the bootstrap, only
 the 95\% confidence intervals are presented since
 the bootstrap estimates were not normally distributed for many of the parameter estimates.
 The results demonstrate a statistically significant positive effect of  biliribin and prothrombin time
  and a negative effect of albumin on transplantation-free survival. However, it should be recognized
that these parameter estimates may be distorted due to the measurement
 error in these longitudinal biomarker measurements.

\begin{table}
\tablewidth=300pt
\caption{The effect of log-transformed biliribin, prothrombin time and
albumin ($p= 1$, 2 and 3, respectively) on transplantation-free survival
using the proposed approach with $M=10$.
We fit model (\protect\ref{1f}) with the true longitudinal
measurements following (\protect\ref{2f})--(\protect\ref{3f}).
     Time-to-event is modeled as a discrete time process with possible event times at 0.5, 1, 2, 3 and 4 years.
 Standard errors (SE) were estimated using the bootstrap. 95\% confidence intervals
 were estimated using the bootstrap with the percentile method (500 bootstrap samples)}\label{t3}
\begin{tabular*}{300pt}{@{\extracolsep{\fill}}ld{3.3}d{3.14}@{}}
\hline
\textbf{Parameters}&\multicolumn{1}{c}{\textbf{Estimate}}&\multicolumn{1}{c@{}}{\textbf{95\% confidence interval}}\\
\hline
$\alpha_{02}$&-5.22&-20.57\mbox{ to }{-}25.06\\
$\alpha_{03}$&-5.48&-21.06\mbox{ to }24.83\\
$\alpha_{04}$&-4.05&-18.67\mbox{ to }27.75\\
$\alpha_{05}$&-4.24&-17.88\mbox{ to }27.45\\
$\alpha_{06}$&-4.48&-18.05\mbox{ to }27.23\\
log Biliribin&0.34&0.02\mbox{ to }1.71\\
log Albumin&-11.39&-61.38\mbox{ to }{-}5.21\\
log Proth&6.16&-3.78\mbox{ to }22.83\\
$\beta_{10}$&0.50&0.22\mbox{ to }0.58\\
$\beta_{11}$&0.26&0.09\mbox{ to }0.35\\
$\beta_{20}$&1.26&0.56\mbox{ to }1.27\\
$\beta_{21}$&-0.05&-0.05\mbox{ to }{-}0.02\\
$\beta_{30}$&2.36&1.06\mbox{ to }2.38\\
$\beta_{31}$&0.02&0.1\mbox{ to }0.03\\
$\sigma_{b10}^2$&0.99&0.45\mbox{ to }1.11\\
$\sigma_{b11}^2$&0.27&0.05\mbox{ to }0.63\\
$\sigma_{b20}^2$&0.03&0.01\mbox{ to }0.04\\
$\sigma_{b30}^2$&0.01&0.005\mbox{ to }0.025\\
\hline
\end{tabular*}
\end{table}

Using the proposed approach, we initially fit model (\ref{1f})--(\ref{3f})  which
incorporated a random intercept and slope term for each of the three
biomarkers. However, the random effect for slope for prothrombin
time and albumin were estimated as nearly zero. Thus, we re-fit the
model without a random slope effect for these two biomarkers.  Table
\ref{t3} shows parameter estimates from the proposed approach with 95\%
confidence intervals estimated using the bootstrap (as in  the
analysis with the observed biomarkers  presented in Table \ref{t2}, we  do
not present the parameter estimates of the standard errors). Except
for the effect of biliribin, estimates of the other two biomarkers
are substantially larger in magnitude under the proposed approach
than when ignoring measurement error and using the observed data
(Table~\ref{t2}).  This is consistent with the common phenomenon that
ignoring measurement error attenuates parameter estimates. In terms
of inference, the effect of prothrombin time on short-term prognosis
is no longer statistically significant with the proposed approach,
while the effects  of biliribin and albumin on prognosis are
statistically significant with both approaches. In the PBC
 analysis, estimates of $\sigma_{p\varepsilon}^2$ were 0.31, 0.12 and
 0.11 for log-transformed values of biliribin, albumin and
 prothrombin time, respectively. The smaller  absolute values
for parameter estimates of albumin and prothrombin time using the
observed markers (Table \ref{t1})  relative to estimates for these markers
using the proposed approach (Table \ref{t2}) can be attributed to
attenuation due to measurement error,  since, in these cases, the
residual variances are substantially larger than the between-subject
variations.

We also conducted analyses where we adjusted for treatment effect
and age in the discrete-time survival model [$\mathbf{Z}_i$ is
treatment group or age in model (\ref{10f})]. As discussed in Section \ref{s23},
we constrained the parameters $\bolds{\zeta}^*_{pT_i}$ so that they
did not vary with $T_i$ (results were similar when we did not
constrain the parameters). When we adjusted for treatment group
(with treatment group coded as 1 for D-penicillamine and 0 for
placebo) in (\ref{10f}), we estimated the $\alpha$ coefficient
corresponding to treatment as 0.051 (95\% CI: $-0.84$ to 1.14).
 The estimates of other parameters were almost
identical to those presented in Table \ref{3f}. When we adjusted for age in
(\ref{10f}), we estimated the $\alpha$ coefficient corresponding to age as
0.018 (95\% CI: 0.01 to 0.129), where age was scaled in units of a
year. Although age was statistically significant, the effects of log
bilirubin, albumin and prothtime on transplantation-free survival
were similar to those for the unadjusted model. Specifically, the
regression coefficients ($\alpha$ coefficients) corresponding to
three markers are 0.394 (95\% CI: 0.07 to 2.21), $-10.65$ ($-83.51$ to
$-5.52$) and 5.76 ($-4.28$ to 30.97).

Table \ref{3f} also shows the estimated fixed effect intercept and slope
for the three longitudinal biomarkers for the proposed approach. The
estimates suggest that biliribin is increasing, while albumin and
prothrombin time are nearly constant over time.

The joint modeling approach is important in this application for a
number of reasons. First, survival models which use observed
biomarkers can result in attenuated estimates of risk. The proposed
approach allowed us to account for the measurement error in
investigating the effect of multiple  biomarker measurements on the
short-term prognosis of PBC patients in terms of
transplantation-free survival. With the proposed approach, we found
that  the ``true'' biliribin and albumin values at the beginning of
an interval  had a sizable and statistically significant effect on
the probability of either a transplantation or death in the
subsequent interval. Second, the proposed approach allows us to
appropriately make inference about changes in the three ``true''
biomarkers over time. As stated before, the largest change over time
was in biliribin which sizably increased over time. When making
these longitudinal inferences, not appropriately modeling the
relationship between the multiple biomarkers and survival may lead
to bias due to informative dropout [\citet{wu}].

\section{Discussion}\label{s5}
We proposed an approach for jointly modeling multivariate
longitudinal  and discrete time-to-event data. Unlike
likelihood-based approaches which require high-dimensional
integration to evaluate the joint likelihood, this approach only
requires fitting bivariate random effects models. This methodology
uses recent methodology for fitting multivariate  longitudinal data
with bivariate linear mixed models proposed by Fieuws et al.
(\citeyear{fieuws2005}, \citeyear{fieuws2007}). They discussed the simple averaging of duplicate
parameters estimates as we did in implementing the proposed
approach. They also proposed a pseudo-likelihood approach which
involves maximizing the sum of likelihoods from bivariate models
across all ${P\choose 2}$ combinations of pairwise longitudinal
markers. Although this later approach may provide some minor
efficiency gain over simple averaging, it would be substantially
more complicated to implement in our setting. Further, one of the
advantages of the pseudo-likelihood approach is that it provides an
analytic expression for the asymptotic variance of the parameter
estimates. Unfortunately, this asymptotic variance is not
generalizable to the joint model with time-to-event data, making the
pseudo-likelihood approach less attractive in our setting.

There are similarities between our approach and the recent approach by Fieuws et al. (\citeyear{fieuws2008}) for predicting
renal graft failure based on multivariate longitudinal profiles. Both approaches model the conditional distribution
of the multivariate longitudinal profiles given the failure time. However, there are major differences between the
two approaches. Fieuws et al. model  the  conditional distribution of the longitudinal measurements given the failure
time and then use  Bayes rule to estimate the probability of failure given the longitudinal profiles.  In our approach, we use
an approximation of the conditional distribution of $\mathbf{X}_i|T_i$ under  the joint model of the multivariate longitudinal and
 time-to-event data in order estimate the parameters of this joint model.

We demonstrated the feasibility of the proposed approach with three biomarkers ($P=3$). However, the approach can
 easily accommodate a large number of longitudinal profiles
  since it simply involves fitting ${P \choose 2}$  bivariate models.
The relationship between the multivariate longitudinal and event-time data is governed by expression (\ref{1f}). However, other functional
relationships are possible with this approach. For example, we could
relate the two processes by averages of ``true'' longitudinal
biomarkers either across time or across different biomarkers.
Alternatively,
 the approach could be formulated so that  the event-time process depends on the  individual's intercept and slope for each of the $P$
 longitudinal biomarkers.

The multivariate longitudinal profiles are modeled as multivariate linear mixed models in (\ref{2f}) and (\ref{3f}), which was appropriate
 for the analysis of the  PBC data. However, the methodology could be extended to allow for more flexible nonlinear modeling of marker profiles.
This would involve approximating the conditional probability
$\mathbf{X}_i|T_i$ in (\ref{5f}) where $h(\mathbf{X}_i|\bolds{\gamma}_i)$
follows  a multivariate nonlinear mixed model, rather than the
linear mixed model discussed in our paper. In the nonlinear case, we
could approximate (\ref{5f}) by (\ref{6f}), where (\ref{6f}) would be a nonlinear mixed
model with parameters indexed by $T_i$ rather than the linear mixed
model presented. However, unless there is biological rational for a
particular nonlinear mixed model, it may be difficult to choose a
reasonable model in most practical situations.

In our formulation, we assumed that event times are only administratively censored after a fixed follow-up at the end of the study.
 For the situation in which patients are censored prematurely, dropout times can be imputed based on a model fit using
patients who had the potential to be followed over the entire study duration.

In this article methodology was developed using a discrete-time
survival model with calibration error being incorporated by using a
probit link function.  This approach led to an analytically
tractable form for incorporating calibration error (\ref{8f}). For the PBC
study, little is lost by using a discrete-time model since the
probability of an event in each of the five intervals is low (the
estimated probability of an event during each of the five time
intervals is 0.05, 0.03, 0.08, 0.07 and 0.07). For continuous-time
survival models such as the Cox model, incorporating calibration is
more difficult since there is no simple analytic solution. That
said,
 we could use a Cox  model if we do not account for the calibration error in replacing
the random effect by their empirical Bayes estimators. Although not
the case in the PBC study, in situations where the within-subject
variation is small relative to the between-subject sources of
variation, the calibration error will be small and there will be
only a small amount of bias induced by not accounting for the
calibration error.

\section*{Acknowledgments}

We thank the Center for Information Technology, NIH, for
providing access to the high-performance computational capabilities
of the Biowulf cluster computer system. We thank the Editor,
Associate Editor and reviewer for their constructive comments which
led to an improved manuscript.

\printaddresses


\begin{thebibliography}{}
\bibitem[\protect\citeauthoryear{Abramowitz and Stegun}{1974}]{abramowitz}
\textsc{Abramowitz, M.} and \textsc{Stegun, I.} (1974). \textit{Handbook of Mathematical Functions}.
Dover, New York.

\bibitem[\protect\citeauthoryear{Albert and Shih}{2009}]{albert}
\textsc{Albert, P. S.} and \textsc{Shih, J. H.} (2009). On estimating the relationship
between longitudinal measurements and time-to-event data using regression
calibration. \textit{Biometrics} DOI:
\href{http://dx.doi.org/10.1111/j.1541-0420.2009.01324.x}{10.1111/j.1541-0420.2009.01324.x}.

\bibitem[\protect\citeauthoryear{Allen et~al.}{2007}]{allen}
\textsc{Allen, C., Duffy., S., Teknos, T. Islam, M., Chen, Z., Albert, P. S.,
Wolf, G. Y.} and \textsc{Van Waes, C.} (2007). A prospective study of serial measurements of NF-$\alpha\beta$ related serum cytokines as biomarkers of response and survival in patients with advanced oropharyngeal squamous cell carcinoma receiving chemoradiation therapy. \textit{Clinical Cancer Research} \textbf{13} 3182--3190.

\bibitem[\protect\citeauthoryear{Chi and Ibrahim}{2006}]{chi}
\textsc{Chi, Y. Y.} and \textsc{Ibrahim, J. G.} (2006). Joint models for multivariate longitudinal and multivariate survival data. \textit{Biometrics} \textbf{62} 432--445.
\MR{2227491}

\bibitem[\protect\citeauthoryear{Brown, Ibrahim and DeGruttola}{2005}]{brown}
\textsc{Brown, E. R., Ibrahim, J. G.} and \textsc{DeGruttola, V.} (2005).
A flexible B-spline model for multiple longitudinal biomarkers and
survival. \textit{Biometrics} \textbf{61} 64--73.
\MR{2129202}

\bibitem[\protect\citeauthoryear{Doran and Lockwood}{2006}]{doran}
\textsc{Doran, H. C.} and \textsc{Lockwood, J. R.} (2006). Fitting value-added models in R.
\textit{Journal of Educational and Behavioral Statistics} \textbf{31} 205--230.

\bibitem[\protect\citeauthoryear{Efron and Tibshirani}{1993}]{efron}
\textsc{Efron, B.} and \textsc{Tibshirani, R. J.} (1993). \textit{An Introduction to the
Boostrap}. Chapman and Hall, New York.
\MR{1270903}

\bibitem[\protect\citeauthoryear{Fieuws and Verbeke}{2005}]{fieuws2005}
\textsc{Fieuws, S.} and \textsc{Verbeke, G.} (2005). Pairwise fitting of mixed models for the joint modelling of  multivariate longitudinal profiles. \textit{Biometrics} \textbf{62} 424--431.
\MR{2227490}

\bibitem[\protect\citeauthoryear{Fieuws, Verbeke and Molenberghs}{2007}]{fieuws2007}
\textsc{Fieuws, S., Verbeke, G.} and \textsc{Molenberghs, G.} (2007). Random-effects models for multivariate repeated measures. \textit{Stat. Methods Med. Res.} \textbf{16} 387--397.
\MR{2405477}

\bibitem[\protect\citeauthoryear{Fieuws et~al.}{2008}]{fieuws2008}
\textsc{Fieuws, S., Verbeke, G., Maes, B.} and \textsc{Vanrenterghem, Y.} (2008). Predicting renal  graft failure using multivariate longitudinal profiles.
\textit{Biostatistics} \textbf{9} 419--431.

\bibitem[\protect\citeauthoryear{Henderson, Diggle and Dobson}{2000}]{henderson}
\textsc{Henderson, R. Diggle, P.} and \textsc{Dobson, A.} (2000). Joint modeling of measurements and event time data. \textit{Biostatistics} \textbf{1} 465--480.

\bibitem[\protect\citeauthoryear{Huang et~al.}{2001}]{huang}
\textsc{Huang, W. H., Zeger, S. L., Anthony, J. C.} and \textsc{Garrett, E.} (2001).
Latent variable model for joint anlaysis of multiple repeated measures and bivariate event times. \textit{Amer. Statist. Assoc.} \textbf{96} 906--914.
\MR{1946363}

\bibitem[\protect\citeauthoryear{Ibrahim, Chen and Sinha}{2004}]{ibrahim}
\textsc{Ibrahim, J. G., Chen, M.} and \textsc{Sinha, D.} (2004). Bayesian methods for jointly modeling of
longitudinal and survival data with applications to cancer vaccine trials. \textit{Statist. Sinica} \textbf{14} 863--883.
\MR{2087976}

\bibitem[\protect\citeauthoryear{Laird and Ware}{1982}]{laird}
\textsc{Laird, N. M.} and \textsc{Ware, J. H.} (1982). Random-effects models for longitudinal  data. \textit{Biometrics} \textbf{38} 963--974.

\bibitem[\protect\citeauthoryear{Murtaugh et~al.}{1994}]{murtaugh}
\textsc{Murtaugh, P. A., Dickson, E. R., Van Dam, G. M., Malincho, M.,
Grambsch, P. M., Langworthy, A. L.} and \textsc{Gips, C. H.} (1994). Primary billary cirrhosis: Prediction
of short-term survival based on repeated patient visits. \textit{Hepatology} \textbf{20} 126--134.

\bibitem[\protect\citeauthoryear{Song, Davidian and Tsiatis}{2002}]{song}
\textsc{Song, X., Davidian, M.} and \textsc{Tsiatis, A. S.} (2002). An estimator
for the proportional hazards model with multiple longitudinal
covariates measured with error. \textit{Biostatistics} \textbf{3} 511--524.

\bibitem[\protect\citeauthoryear{Tsiatis and Davidian}{2004}]{tsiatis2004}
\textsc{Tsiatis, A. A.} and \textsc{Davidian, M.} (2004). Joint modeling of longitudinal and time-to-event data:
An overview. \textit{Statist. Sinica} \textbf{14} 809--834.
\MR{2087974}

\bibitem[\protect\citeauthoryear{Tsiatis, DeGruttola and Wulfsohn}{1995}]{tsiatis1995}
\textsc{Tsiatis, A. A., DeGruttola, V.} and \textsc{Wulfsohn, M. S.} (1995).
Modeling the relationship of survival to longitudinal data measured with error. Applications to survival and CD4 counts in patients with AIDS.
\textit{J. Amer. Statist. Assoc.} \textbf{90} 27--37.

\bibitem[\protect\citeauthoryear{Venables, Smith and the R Development Core Team}{2008}]{venables}
\textsc{Venables, W. N., Smith, D. M.} and \textsc{the R Development Core Team} (2008).
An Introduction to R. Version 2.8.1 (2008-12-22).

\bibitem[\protect\citeauthoryear{Verbeke and Molenberghs}{2000}]{verbeke}
\textsc{Verbeke, G.} and \textsc{Molenberghs, G.} (2000). \textit{Linear Mixed Models for Longitudinal Data}. Springer, New York.
\MR{1880596}

\bibitem[\protect\citeauthoryear{Wei and Tanner}{1990}]{wei}
    \textsc{Wei, G. C. G.} and \textsc{Tanner, M. A.} (1990).
    A Monte-Carlo implementation of the E--M algorithm and the poor man's data
    augmentation algorithm. \textit{J. Amer. Statist. Assoc.}
    \textbf{85} 699--704.

\bibitem[\protect\citeauthoryear{Wu and Carroll}{1988}]{wu}
\textsc{Wu, M. C.} and \textsc{Carroll, R. J.} (1988). Estimation and comparison
of changes in the presence of informative right censoring by
modeling the censoring process. \textit{Biometrics} \textbf{45} 939--955.
\MR{0931633}

\bibitem[\protect\citeauthoryear{Wulfsohn and Tsiatis}{1997}]{wulfsohn1997}
\textsc{Wulfsohn, M. S.} and \textsc{Tsiatis, A. A.} (1997). A joint model for survival and longitudinal data measured with error. \textit{Biometrics} \textbf{53} 330--339.
\MR{1450186}

\bibitem[\protect\citeauthoryear{Xu and Zeger}{2001a}]{xu2001a}
\textsc{Xu, J.} and \textsc{Zeger, S. L.} (2001a). Joint analysis of longitudinal data comprising repeated measures and times to events. \textit{Appl. Statist.} \textbf{50} 375--387.
\MR{1856332}

\bibitem[\protect\citeauthoryear{Xu and Zeger}{2001b}]{xu2001b}
\textsc{Xu, J.} and \textsc{Zeger S. L.} (2001b). The evaluation of multiple
surrogate endpoints. \textit{Biometrics} \textbf{57} 81--87.
\MR{1833292}

\end{thebibliography}
\end{document}